\documentclass{ws-ijmpc}

\begin{document}

\markboth{Nuno Crokidakis}
{Dynamics of drug trafficking: Results from a simple compartmental model}

\catchline{}{}{}{}{}

\title{Dynamics of drug trafficking: Results from a simple compartmental model}


\author{Nuno Crokidakis $^{*}$}

\address{
Instituto de F\'{\i}sica, \hspace{1mm} Universidade Federal Fluminense \\
 Niter\'oi - Rio de Janeiro, \hspace{1mm} Brazil \\ 
$^{*}$ nunocrokidakis@id.uff.br}

\maketitle

\begin{history}
\received{Day Month Year}
\revised{Day Month Year}
\end{history}

\begin{abstract}
\noindent
In this work we propose a simple model for the emergence of drug dealers. For this purpose, we built a compartmental model considering four subpopulations, namely susceptibles, passive supporters, drug dealers and arrested drug dealers. The target is to study the influence of the passive supporters on the long-time prevalence of drug dealers. Passive supporters are people who are passively consenting to the drug trafficking cause. First we consider the model on a fully-connected newtork, in such a way that we can write a rate equation for each subpopulation. Our analytical and numerical results show that the emergence of drug dealers is a consequence of the rapid increase number of passive supporters. Such increase is associated with a nonequilibrium active-absorbing phase transition. After that, we consider the model on a two-dimensional square lattice, in order to compare the results in the presence of a simple social network with the previous results. The Monte Carlo simulation results suggest a similar behavior in comparison with the fully-connected network case, but the location of the critical point of the transition is distinct, due to the neighbors' correlations introduced by the presence of the lattice.


\end{abstract}

\ccode{PACS Nos.: 05.10.-a, 05.70.Jk, 87.23.Ge, 89.75.Fb}

\section{Introduction}

\qquad The study of criminality has been the subject of interest for sciences like mathematics in the last years (for recent reviews, see  \cite{review,donnay,pacheco,sooknanan}). Usual methods consider the range from partial differential equations and self-exciting point processes to agent-based models, evolutionary game theory and network science \cite{review}. Recent reviews on mathematical models of social contagion and criminality can be found in Refs. \cite{sooknanan,sooknanan2}.

Considering the particular problem of drug trafficking, a  dynamic model was proposed to analyze the proliferation of drugs, based on the presence of two classes of individuals, namely dealers and producers. The model shows that equilibrium with no dealers and producers is fairly difficult. The model predicts a nonzero steady-state stable equilibrium, in which the number of dealers and producers can be kept at low levels if the repression against these activities focuses on their mutually reinforcing interaction. In this case, a reliable policy against drug proliferation should contemplate simultaneously the sides of supply and demand. The possibility of an equilibrium steady state with the total population being users or producers is excluded \cite{araujo}.

In countries like Brazil the drug traffiking is a huge problem. A recent work suggested an increase in incarceration rates in the Brazilian southern state of Rio Grande do Sul during the last decade, especially due to drug trafficking offenses. The authors connected such increase due to two factors: the increase in the number of trafficking and criminal gangs in the state observed after 2005; and the changes in the law and proposed distinct sanctions for users and traffickers \cite{ornell}.

Concerning mathematical analysis of criminality, some authors considered compartmental models in order to study the spreading of crime. The authors in \cite{misra} considered a nonlinear mathematical model  to study the effect of police force in controlling crime in a society with variable population size. The authors show that the model has only one equilibrium, namely crime persistent equilibrium. This equilibrium always exists and is locally as well as globally stable under certain conditions. The authors in \cite{nuno} present a mathematical model of a criminal-prone self-protected society that is divided into socio-economical classes. They studied the effect of a non-null crime rate on a free-of-criminals society which is taken as a reference system. A relevant conclusion that can be derived from the study is that the kind of systems under consideration are criminal-prone, in the sense that criminal-free steady states are unstable under small perturbations in the socio-economical context \cite{nuno}. The authors in \cite{mataru} discussed that not all kind of crimes can be eradicated. Thus, they proposed a compartmental model in order to study the eradication of unemployment-related crimes in the developing countries. The results suggest that vocational training  and employment strategies are more effective in combating crime when applied simultaneously \cite{mataru}. Another compartmental model incorporated education programs as tools to assess the population-level impact on the spread of crime. With no compliance, the authors observed a high level of active criminal population, and if the compliance rates are significantly improved, the active population level decreases \cite{kwofie}. The impact of legal and illegal guns on the growth of violent crimes was also analyzed through mathematical models in recent papers \cite{monteiro,meu}.

Another work considered a two-dimensional lattice model for residential burglary, where each site is characterized by a dynamic attractiveness variable, and where each criminal is represented as a random walker \cite{short}. The authors considered that the dynamics of criminals and of the attractiveness field are coupled to each other via specific biasing and feedback mechanisms. They concluded that, depending on parameter choices, several regimes of aggregation, including hotspots of high criminal activity, can be described by the simple model. One can also mention a work where the authors considered a model based on the predator-prey problem, in order to study the interaction between criminal population and non-criminal population. Considering a law enforcement term in the model's equations, the authors discussed that the criminal minded population exist as long as coefficient of enforcement does not cross a threshold value and after this value the criminal minded population extinct \cite{abbas}.

An important concept concerning criminal activities is the idea of passive supporters. The passive supporters were introduced in mathematical models in the context of spreading of terrorism \cite{galam_epjb_2002}. The passive supporters do not oppose a terrorist act. They go unnoticeable and most of them reject the violent aspect of the terrorist action. They only share in part their cause \cite{galam_2003,galam_2003_2,galam_2023}. The mentioned works \cite{galam_epjb_2002,galam_2003,galam_2003_2} showed that the presence of such individuals, passive supporters, are a key feature to understand the spreading of terrorism. Taking this idea in mind, we consider the presence of passive supporters in the dynamics of drug trafficking. We will discuss in this manuscript how even a small fraction of such passive supporters can be effective in the emergence of drug dealers in a population where they do not exist at the beginning. For this purpose, we built a mathematical compartmental model consisting of four subpopulations. Our analytical and numerical results show that the emergence of drug dealers is consequence of the rapid increase number of passive supporters. Such increase can be associated with a nonequilibrium phase transition in the language of Physics of critical phenomena \cite{marro2005nonequilibrium}.


\section{Model}

\qquad Let us consider a populations of $N$ agents or individuals. Each individual can be in one of four possible states or compartments, namely: (1) susceptible individual (\textbf{S}), that is an individual that was never a drug dealer or he/she was in the past and quit; (2) passive supporter of drug trafficking (\textbf{P}), an agent that is not a drug dealer but he/she is a person who is passively consenting to the drug trafficking cause. This can be occur, for example, due to the perception that if the government allows the drug trafficking, less people will die in the ``war'' (police x drug dealers) \footnote{The mentioned war has been the cause of a lot of deaths in cities where the drug trafficking is a great social problem, like Rio de Janeiro, Brazil \cite{jstor}.}; (3) drug dealer (\textbf{D}), that is an agent that do the activity of seeling drugs; and (4) arrested drug dealer (\textbf{A}), a drug dealer that was arrested by the police. The following microscopic rules control the dynamics of the population:

\begin{itemize}
\item $S \stackrel{\beta}{\rightarrow} P$: a susceptible agent (\textbf{S}) becomes a passive supporter (\textbf{P}) with probability $\beta$ if he/she is in contact with passive suporters (\textbf{P});

\item $P \stackrel{\delta}{\rightarrow} D$: a passive supporter (\textbf{P}) becomes a drug dealer (\textbf{D}) with probability $\delta$ if he/she is in contact with drug dealers (\textbf{D});

\item $P \stackrel{\sigma}{\rightarrow} D$: a passive supporter (\textbf{P}) can also becomes a drug dealer spontaneously, by his/her own free will, with probability $\sigma$;

\item $D \stackrel{\gamma}{\rightarrow} A$: a drug dealer (\textbf{D}) can be arrested by police and he/she becomes an arrested drug dealer (\textbf{A}) with probability $\gamma$;

\item $A \stackrel{\alpha}{\rightarrow} D$: an arrested drug dealer (\textbf{A}), after leave the prison, can come back to the criminal activity (drug trafficking), i.e., he/she becomes again a drug dealer (\textbf{D}) with probability $\alpha$;

\item $A \stackrel{1-\alpha}{\rightarrow} S$: an arrested drug dealer (\textbf{A}), after leave the prison, can also abandon the criminal activity and he/she becomes a susceptible agent (\textbf{S}) with the complementary probability $1-\alpha$;

\end{itemize}

Let us discuss briefly about the transitions probabilities. The probability $\beta$ is a contagion probability, it quantifies the influence of passive suporters over susceptible individuals. Passive supporters do not act as drug dealers, but they are not against the criminal activity. Thus, passive supporters can engaje susceptible agents to become also passive suporters of drug trafficking. The probability $\delta$ also models a social contagion, in such a case the influence of drug dealers over passive supporters. Since passive supporters are favorable to drug trafficking, they can become also drug dealers over the influence of drug dealers with probability $\delta$, or can turn in drug dealers spontaneously with probability $\sigma$. It is not unusual in Brazil that some middle-class individuals become drug dealers due to social contacts with drug dealers. Indeed, two of such individuals, \textit{Pedro Dom} and \textit{Playboy}, have histories that lead to the creation of TV series and movies \cite{hp1,hp2}. As we will show in the sequence, the probability $\sigma$ is a key parameter of the model to keep the long-run survival of drug trafficking. Since drug dealers are practicing a criminal activity, they can be arrested by the police action with probability $\gamma$. An arrested drug dealer, after leaving the prison, can choose one of two possibilities: return to the criminal activity (with probability $\alpha$) or abandon such criminal activities and come back to the susceptible population (probability $1-\alpha$).

In the next section we will discuss our analytical and numerical results.


\section{Results}

\subsection{Fully-connected population}

\qquad If we consider that the $N$ individuals in the population are fully mixed, we can write the mean-field rate equations for the time evolution of the subpopulations of susceptible agents $S(t)$, the passive supporters $P(t)$, the drug dealers $D(t)$ and the arrested drug dealers $A(t)$. Defining the four subpopulation densities, namely $s(t)=S(t)/N, p(t)=P(t)/N, d(t)=D(t)/N$ and $a(t)=A(t)/N$, the rate equations can be written as follows:
\begin{eqnarray} \label{eq7}
\frac{d}{dt}s(t) & = & -\beta\,s(t)\,p(t) + (1-\alpha)\,a(t) ,\\ \label{eq8}
\frac{d}{dt}p(t) & = & \beta\,s(t)\,p(t) - \sigma\,p(t) - \delta\,p(t)\,d(t) , \\ \label{eq9}
\frac{d}{dt}d(t) & = & \sigma\,p(t) + \delta\,p(t)\,d(t) - \gamma\,d(t) + \alpha\,a(t) , \\ \label{eq10}
\frac{d}{dt}a(t) & = & \gamma\,d(t) - a(t) ~.
\end{eqnarray}
\noindent
For simplicity, we considered a fixed population, thus we have the normalization condition,
\begin{equation} \label{eq11}
s(t)+p(t)+d(t)+a(t) = 1 ~,
\end{equation}
\noindent
valid at each time step $t$.

Let us start considering the behavior of the model for short times. As it is usual in contagion epidemic models, we consider as initial condition the introduction of a single ``infected'' individual. In our case this means one passive supporter, i.e., our initial conditions are given by the densities $p(0)=1/N, s(0)=1-1/N$ and $d(0)=a(0)=0$. In such a case, one can linearize Eq. (\ref{eq8}) to obtain \cite{bailey}
\begin{equation} \label{eq12}
\frac{d}{dt}p(t) = (\beta-\sigma)\,p(t)  ~,
\end{equation}  
\noindent
that can be directly integrated to obtain $p(t)=p(0)\,e^{\sigma\,(R_o-1)\,t}$, where $p(0)=p(t=0)$ and one can obtain the basic reproductive number,
\begin{equation} \label{eq13}
R_o = \frac{\beta}{\sigma} ~.
\end{equation}

As it is standard in epidemic models \cite{bailey}, we will see an outbreak and the persistence of the ``disease'' (drug trafficking \footnote{Since the passive supporters are responsible for the emergence of drug dealers, the occurrence of an outbreak in $p(t)$ will lead to the persistence of drug dealers in the long-time limit.}) in the long-time run if $R_o > 1$, i.e., for $\beta > \sigma$. In other words, for the initial time evolution of the population, the contagion probability $\beta$ and the spontaneous transition probability $\sigma$ governs the dynamics. As pointed in section II, the parameter $\sigma$ is a key parameter in the dynamics of the model. Indeed, looking for the above-mentioned initial condition, namely $p(0)=1/N, s(0)=1-1/N$ and $d(0)=a(0)=0$, we see that we start the model with no drug dealers $D$ and no arrested drug dealers $A$. Thus, the transition $P \to D$ will not occur in the initial times due to social contagion among drug dealers and passive supporters. If $\sigma=0$ the spontaneous transition $P\to D$ will not yet occur in the initial times. If the dynamics evolves in time, we wil observe in the long-time run the population in the state $p=1$ and $s=d=a=0$. In the language of Nonequilibrium Statistical Physics, it is an absorbing state since there are only passive supporters ($P$) in the population, and the dynamics becomes frozen since no transitions will occur anymore \cite{marro2005nonequilibrium,hinrichsen2000non}. Such equilibrium solution $(s,p,d,a)=(0,1,0,0)$ for $\sigma=0$ was confirmed through the numerical integration of Eqs. (\ref{eq7}) - (\ref{eq10}) (not shown). In addition, if we look for Eq. (\ref{eq12}) with $\sigma=0$, we can see that the fraction $p(t)$ will grow exponentially in the form $p(t)=p(0)\,e^{\beta\,t}$. Thus, in the following we will consider the model always with $\sigma\neq 0$.

\begin{figure}[t]
\begin{center}
\vspace{6mm}
\includegraphics[width=0.45\textwidth,angle=0]{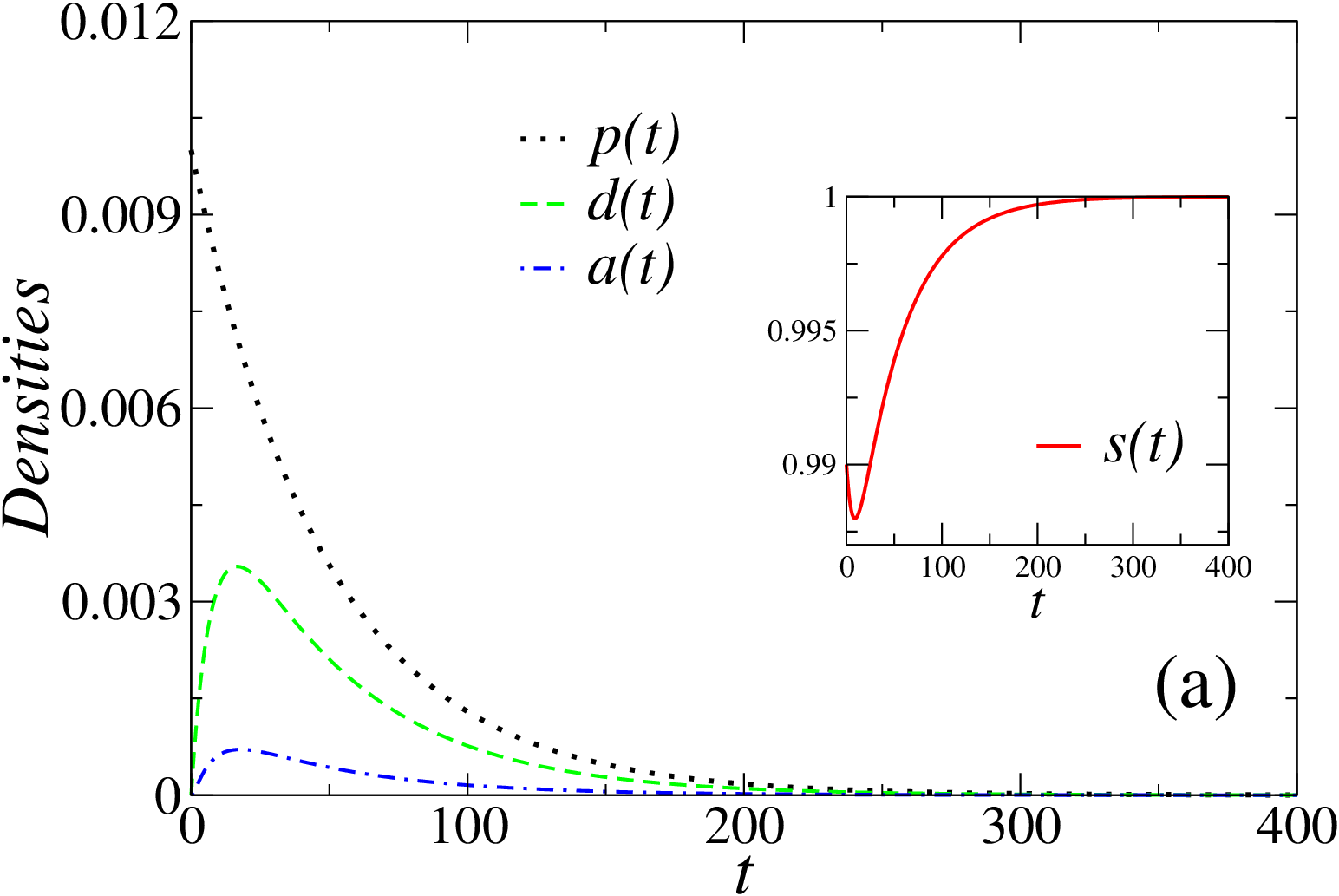}
\hspace{0.3cm}
\includegraphics[width=0.45\textwidth,angle=0]{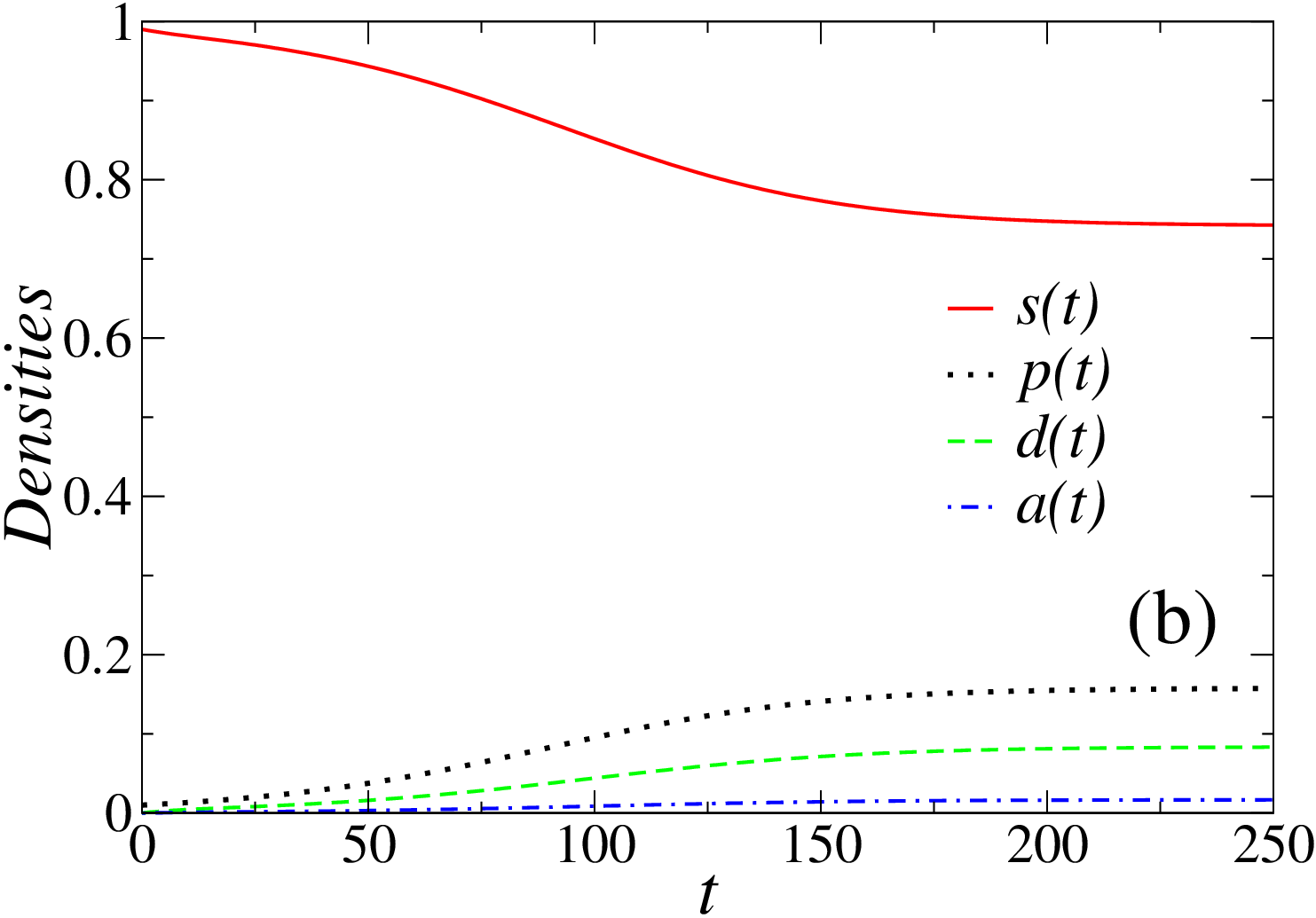}
\\
\vspace{1.0cm}
\includegraphics[width=0.45\textwidth,angle=0]{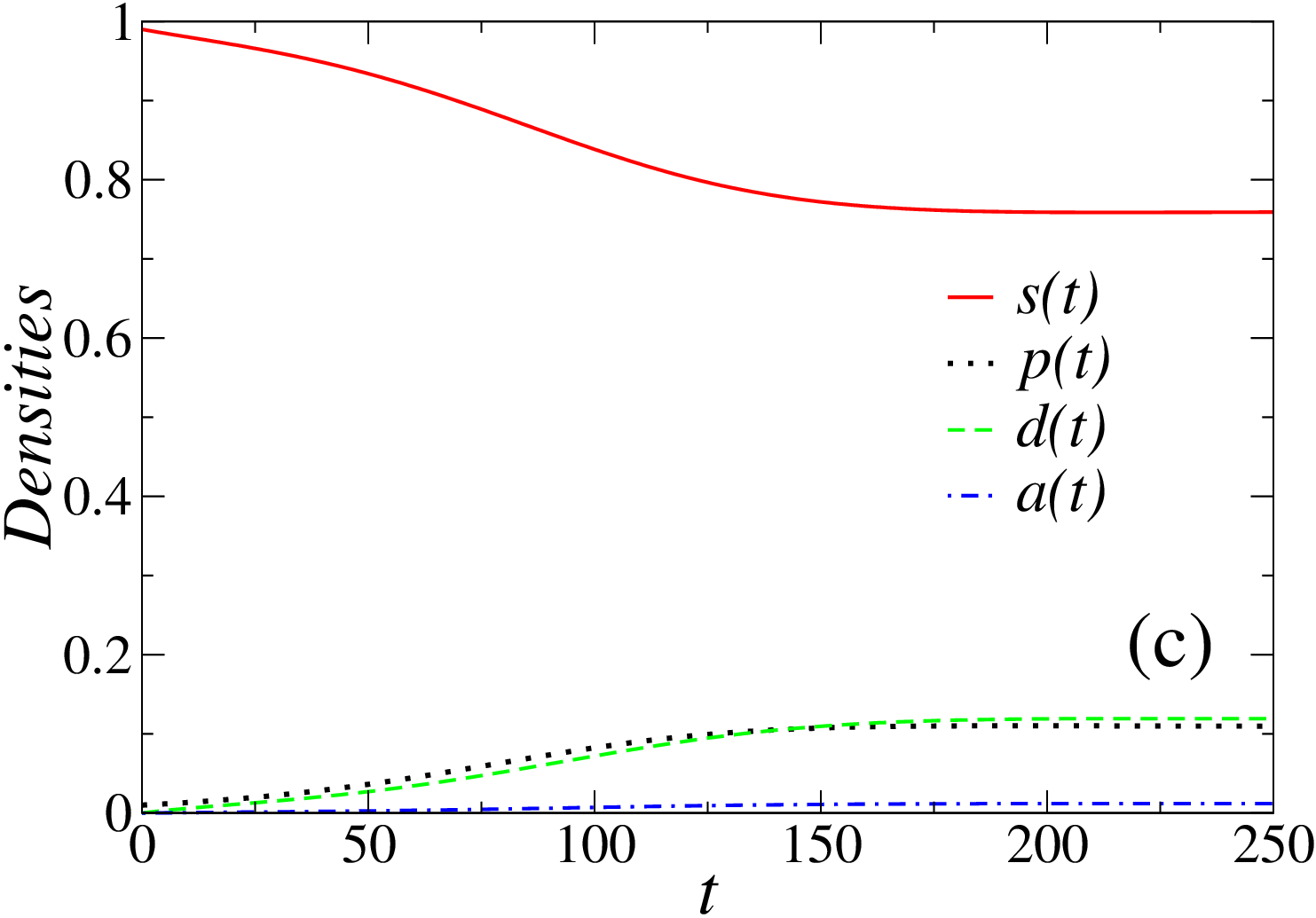}
\hspace{0.3cm}
\includegraphics[width=0.45\textwidth,angle=0]{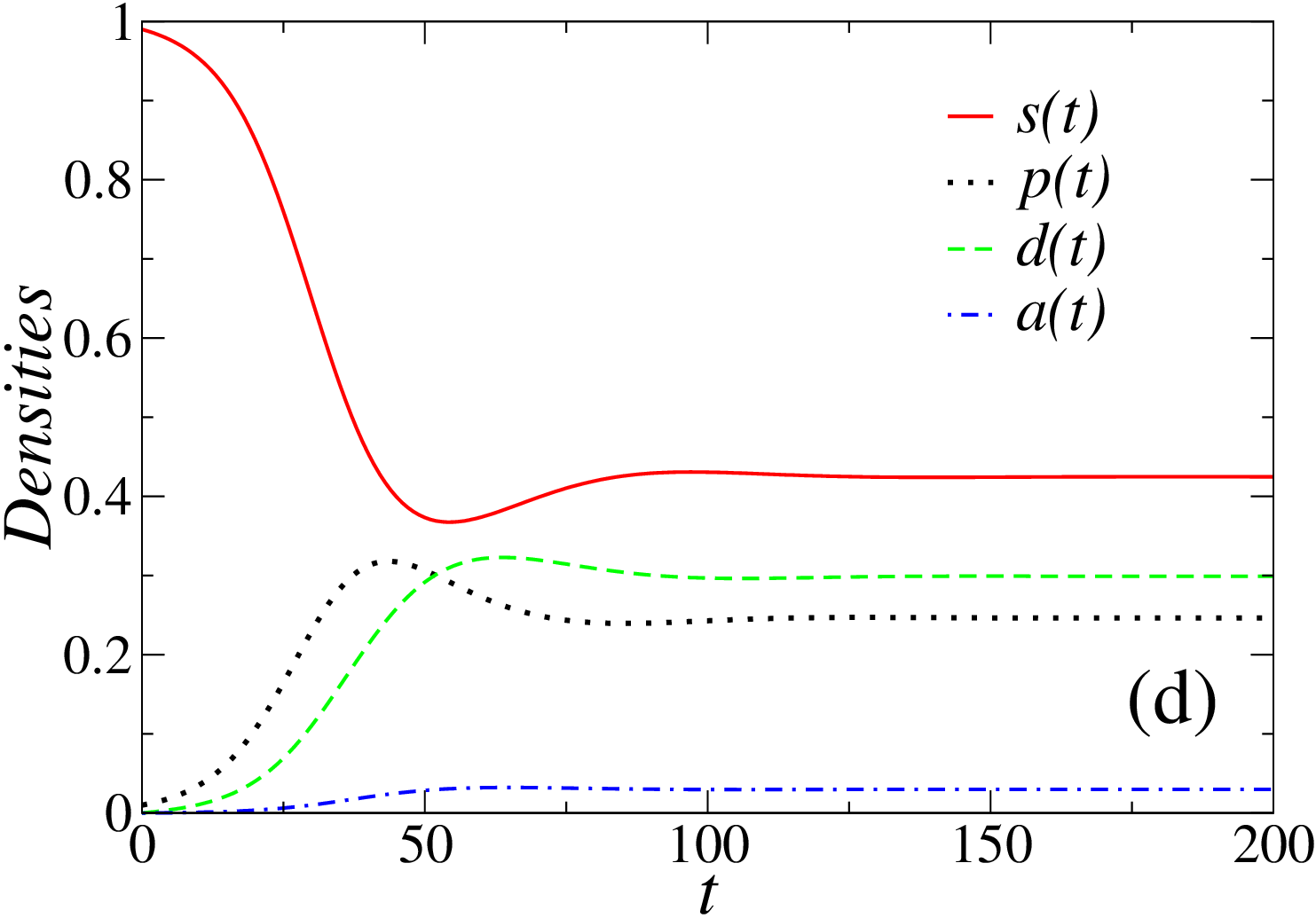}
\end{center}
\caption{Time evolution of the four densities of agents $s(t), p(t), d(t)$ and $a(t)$ for the mean-field formulation of the model, based on the numerical integration of Eqs. (\ref{eq7}) - (\ref{eq10}). The fixed parameters are $\delta=0.05, \alpha=0.30$ and $\sigma=0.07$, and we varied the parameters $\beta$ and $\gamma$: (a) $\beta=0.05, \gamma=0.20$, (b) $\beta=0.10, \gamma=0.20$, (c) $\beta=0.10, \gamma=0.10$, (d) $\beta=0.20, \gamma=0.10$. For such values, we have from Eq. (\ref{eq13}): (a) $R_o\approx 0.71$, (b) $R_o \approx 1.43$, (c) $R_o \approx 1.43$, (d) $R_o \approx 2.86$.}
\label{fig1}
\end{figure}

Now we can analyze the time evolution of the subpopulation densities $s(t), p(t), d(t)$ and $a(t)$. We fixed the parameters $\delta=0.05, \alpha=0.30$ and $\sigma=0.07$ and varied the parameters $\beta$ and $\gamma$. Results are exhibited in Fig. \ref{fig1}. For fixed $\gamma=0.20$, one can see in Fig. \ref{fig1} (a) that a small value of the contagion probability $\beta$ like $\beta=0.05$ leads the population to evolve to a state where the populations $p, d$ and $a$ disapper of the system after a long time, and there will be only suscetible individuals in the population. This represents an absorbing state since the dynamics will become frozen with the extinction of the three subpopulations $p, d$ and $a$. This result will be discussed in details analytically in the following. Keeping $\gamma=0.20$ and incresing $\beta$ to $\beta=0.10$ this mentioned absorbing state will not occur anymore and we can see the coexistence of the four subpopulations (panel (b)). Looking now for fixed $\gamma=0.10$ and increasing the contagion probability $\beta$, we see an increase of the populations $p, d$ and $a$ and a deacrease of $s$ (panels (c) and (d)). Since the contagion probability leading to the transition $S \to P$ increases, we observe an increase of the passive supporters $P$ that leads to an increase of the drug dealers $D$. With more drug dealers, we observe the increase of the number of arrested drug dealers $A$.

Taking the long-time limit $t\to\infty$ to obtain the stationary densities of the model, $s=s(t\to\infty), p=p(t\to\infty), d=d(t\to\infty)$ and $a=a(t\to\infty)$, and considering the density $p$ as the order parameter of the model we can find a phase transition at a critical point (see Appendix)
\begin{eqnarray} \label{eq14}
\beta_{c} = \sigma
\end{eqnarray}
\noindent
As also discussed in the appendix, for $\beta \leq \beta_{c}$ the population is in an absorbing phase where there are only susceptible individuals in the stationary states, which leads to the first solution of the model:
\begin{eqnarray} \label{eq15}
s & = & 1 \\ \label{eq16}
p & = & 0 \\ \label{eq17}
d & = & 0 \\ \label{eq18}
a & = & 0
\end{eqnarray}

\begin{figure}[t]
\begin{center}
\vspace{6mm}\
\includegraphics[width=0.65\textwidth,angle=0]{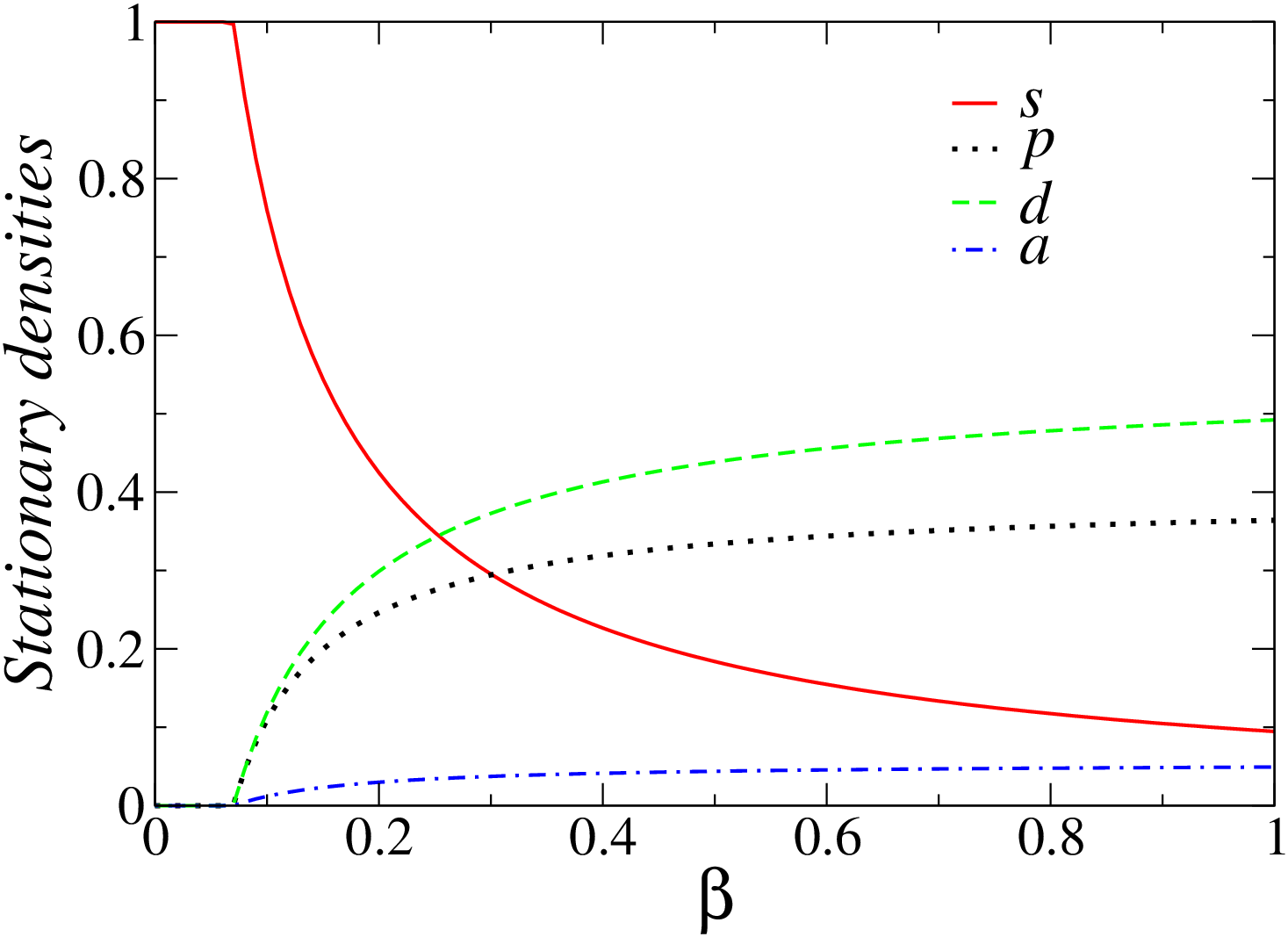}
\end{center}
\caption{Stationary densities $s, p, d$ and $a$ as functions of the contagion probability $\beta$ for $\gamma=0.10, \delta=0.05, \alpha=0.30$ and $\sigma=0.07$. The lines were obtained from Eqs. (\ref{eq15}) - (\ref{eq22}). For the considered parameters, the critical point is given by $\beta_c=0.07$. It is important to note that these behaviors are present also for other parameter values, and what is shown here works as a pattern.}
\label{fig2}
\end{figure}

In addition, for $\beta> \beta_{c}$ the four subpopulations coexist in the steady states, and we have a second solution for the model:
\begin{eqnarray} \label{eq19}
s & = & \frac{\delta\,d+\sigma}{\beta} \\ \label{eq20}
p & = & \frac{(1-\alpha)\,\gamma\,d}{\sigma+\delta\,d} \\ \label{eq21}
a & = & \gamma\,d
\end{eqnarray}
\noindent  
where $d$ is given by
\begin{equation} \label{eq22}
d = \frac{c_2}{2\,c_1}\,\left\{-1 + \sqrt{1-\frac{4\,c_1\,c_3}{c_2^{2}}}\right\} 
\end{equation}
\noindent
and the coefficients $c_1, c_2$ and $c_3$ are given by (see the Appendix)
\begin{eqnarray} 
c_1 & = & \delta\,[\delta+(1+\gamma)\,\beta] \\ 
c_2 & = &  2\,\sigma\,\delta + [(1-\alpha)\,\gamma+(1+\gamma)\,\sigma-\delta]\,\beta \\ 
c_3 & = & \sigma\,(\sigma-\beta)
\end{eqnarray}

An illustration of the long-time behavior of the population is exhibited in Fig. \ref{fig2}, where we plot the stationary densities $s, p, d$ and $a$ as functions of the contagion probability $\beta$ for fixed parameters $\gamma=0.10, \delta=0.05, \alpha=0.30$ and $\sigma=0.07$. For such values, we have $\beta_c=0.07$. The lines were obtained from the numerical integration of Eqs. (\ref{eq7}) - (\ref{eq10}), and agree very well with the analytical solutions: for $\beta<\beta_c$ we have the solution given by Eqs. (\ref{eq15}) - (\ref{eq18}), and for $\beta>\beta_c$ the solution is given by Eqs. (\ref{eq19}) - (\ref{eq22}). The increase of the contagion probability $\beta$ leads to an increase of the passive supporter subpopulation. Since such passive supporters are agents that are not drug dealers but they are persons who are passively consenting to the drug trafficking cause, they are susceptible to the social influence of drug dealers (see the social interaction $P+D \to D + D$). Thus, the increase of the passive supporters leads to the increase of the population of drug dealers in the stationary states, as we see in Fig. \ref{fig2}. For the considered parameters in Fig. \ref{fig2}, one can see that for $\beta > \approx 0.26$ the stationary population of drug dealers becomes the majority subpopulation in the system. Also, we observe that the population of susceptible agents does not disappear in the steady state, even for the limiting case $\beta=1.0$.

\begin{figure}[t]
\begin{center}
\vspace{6mm}
\includegraphics[width=0.45\textwidth,angle=0]{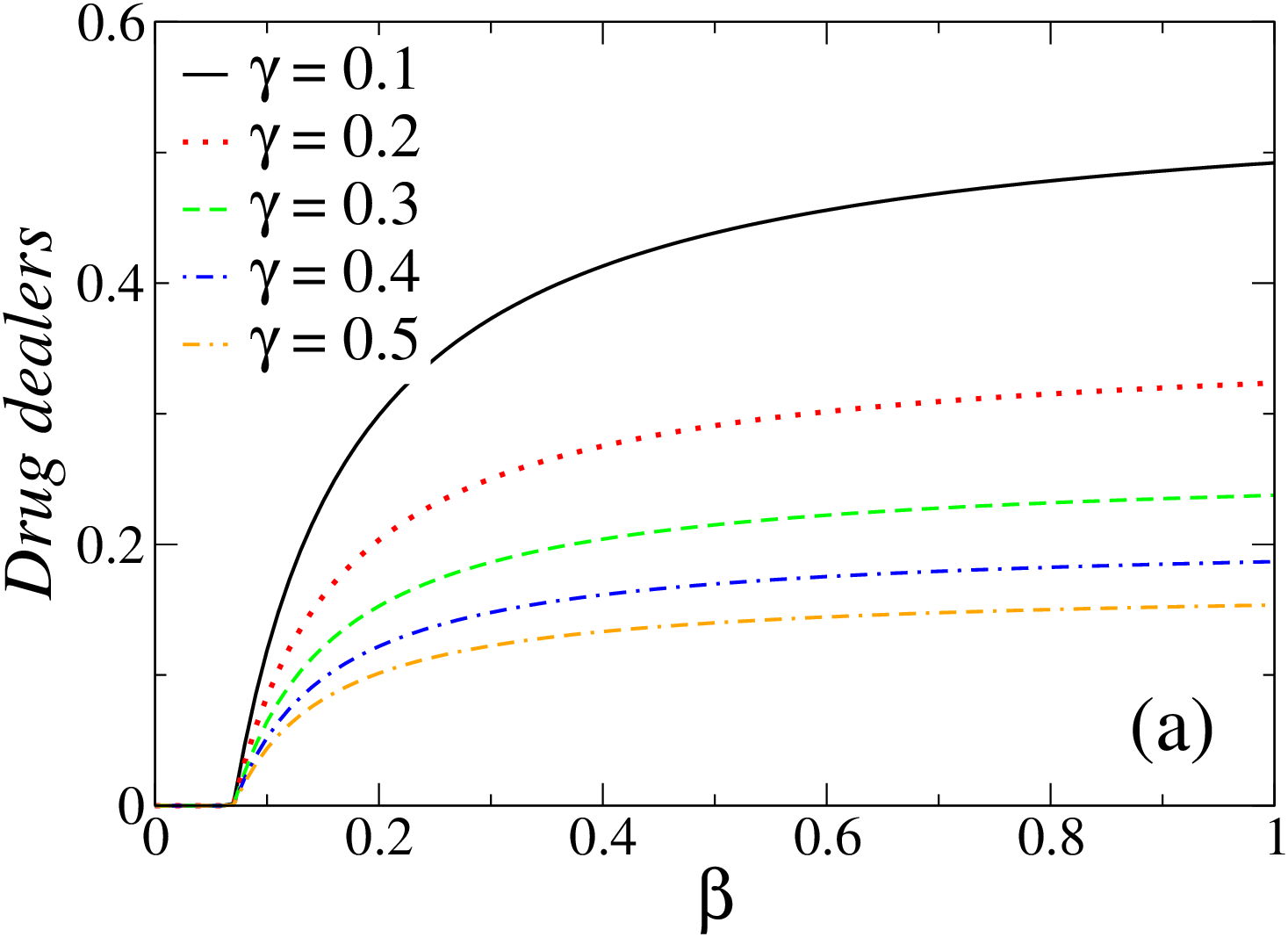}
\hspace{0.3cm}
\includegraphics[width=0.45\textwidth,angle=0]{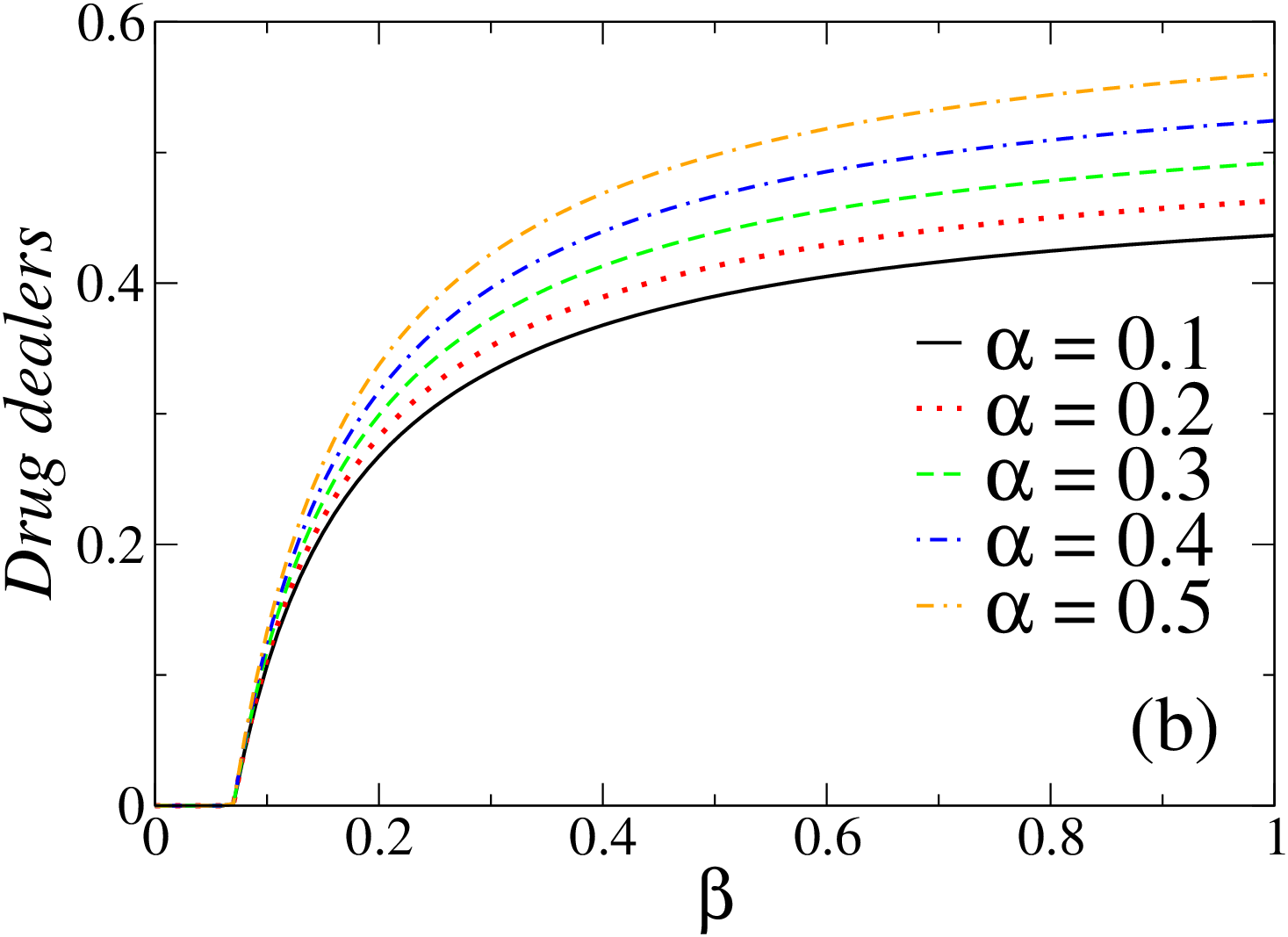}
\\
\vspace{1.0cm}
\includegraphics[width=0.45\textwidth,angle=0]{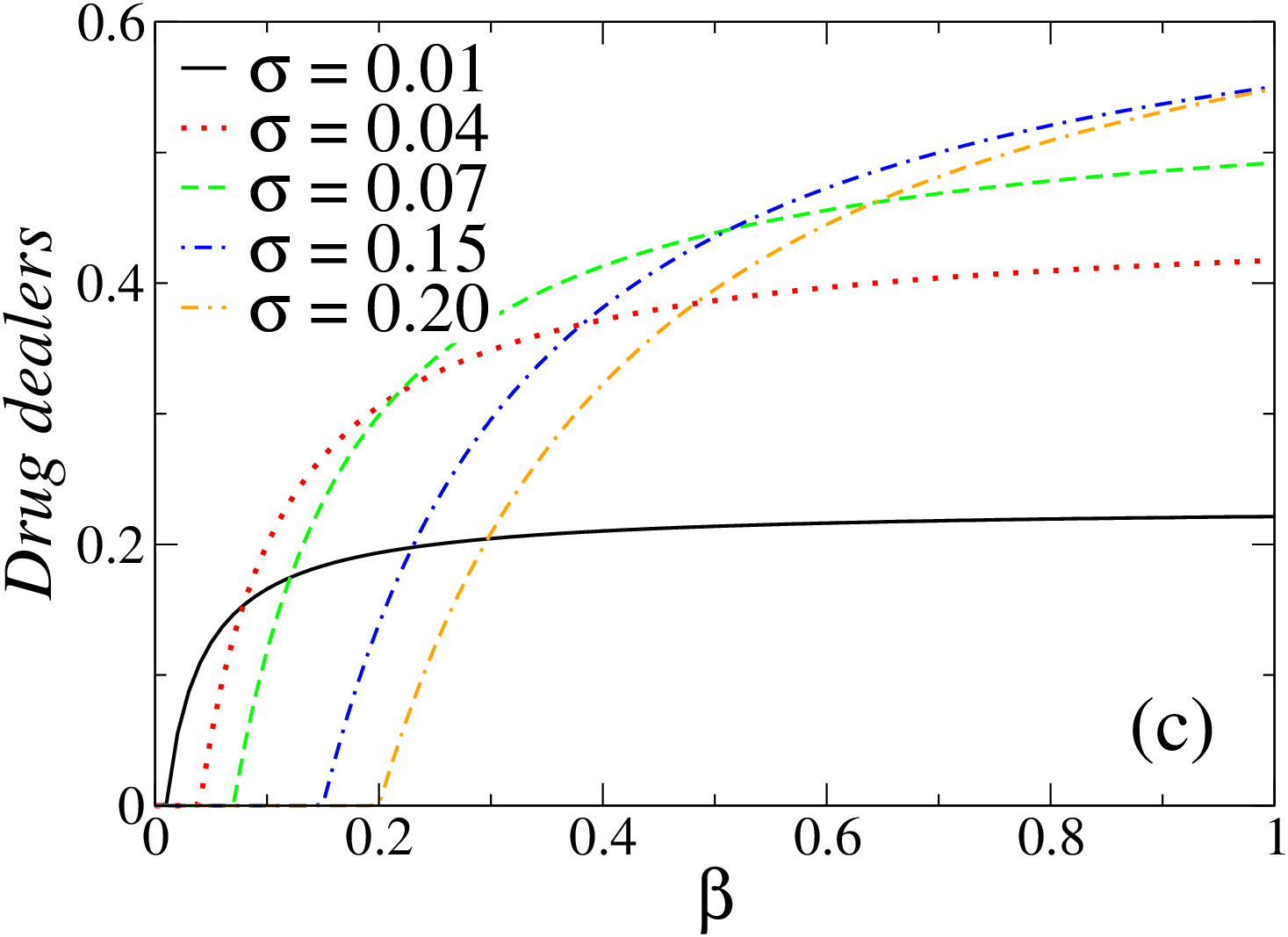}
\end{center}
\caption{Stationary density of drug dealers $d$ as functions of the contagion probability $\beta$. We exhibit in the distinct panels the variation of distinct parameters, for fixed $\delta=0.05$: (a) fixed $\alpha=0.3, \sigma=0.07$ and typical values of the arrest probability $\gamma$; (b) fixed $\gamma=0.1, \sigma=0.07$ and typical values of the probability to come back to crime after a drug dealer is released $\alpha$; (c) $\gamma=0.1, \alpha=0.3$ and typical values of the spontaneous transition $P\to D$ probability $\sigma$. Notice that, as discussed in the text, the critical point $\beta_c$, indeed, does not depend on $\gamma$ and $\alpha$, as it can be seen in panels (a) and (b), but it depends on $\sigma$, as exhibited in panel (c).}
\label{fig3}
\end{figure}

In order to analyze the effects of the parameters on the population of drug dealers, we exhibit in Fig. \ref{fig3} the stationary values of the density of drug dealers as functions of $\beta$, for fixed parameter $\delta=0.05$. The lines were obtained from the numerical integration of Eqs. (\ref{eq7}) - (\ref{eq10}). In Fig. \ref{fig3} (a) we fixed $\alpha=0.3, \sigma=0.07$ and the arrest probability $\gamma$ is varied. We can see that the increase of such probability can effectively decrease the stationary values of drug dealers. Considering for example $\beta=0.3$, we have $d\approx 0.373$ for $\gamma=0.1, d=0.186$ for $\gamma=0.3$ and $d=0.122$ for $\gamma=0.5$. On the other hand, in Fig. \ref{fig3} (b) we fixed $\gamma=0.1, \sigma=0.07$ and the probability to come back to crime after a drug dealer is released $\alpha$ is varied. We can see that, in such a case, the values of $d$ does not change rapidly as in the previous case, where we varied $\gamma$. Looking again for fixed $\beta=0.3$, we have $d\approx 0.422$ for $\alpha=0.5, d\approx 0.373$ for $\alpha=0.3$ and $d\approx 0.332$ for $\alpha=0.1$. However, the values of $d$ increases when we increaase $\alpha$, i.e., if the majority of drug dealers come back to their criminal activities after leaving the prison, the number of drug dealers tend to increase. Looking for references \cite{brazil,usa}, we can see that $70\%$ of prisoners released in 2012 in USA were arrested again within five years, according to data from the Bureau of Justice Statistics (BJS) \cite{usa}. The recidivism rate is over $70 - 80\%$ for prisoners with juvenile records. Similar percentages are observed in Brazil \cite{brazil}. Finally, another illustration of the key importance of the parameter $\sigma$ in the model, the responsible for the spontaneous transition $P\to D$, is shown in Fig. \ref{fig3} (c), where we fixed $\gamma=0.1$ and $\alpha=0.3$. We can see that this parameter determines the position of the critical point $\beta_c$, as predicted in Eq. (\ref{eq14}), as well as it leads the number of drug dealers to increase rapidly.


\subsection{Two-dimensional square lattice}

\qquad In this subsection we present some numerical results of numerical Monte Carlo simulations of the model on two-dimensional grids (square lattices). Thus, we built an agent-based formulation of the compartmental model proposed in the last subsection. For this purpose, the algorithm to simulate the model is defined as follows:

\begin{itemize}

\item we generate a $L$ x $L$ grid or square lattice with a population size $N=L^{2}$ and periodic boundary conditions;

\item given an initial condition $S(0), P(0), D(0)$ and $A(0)$, we randomly distribute these agents in the lattice sites;

\item at each time step, every lattice site is visited in a sequential order;

\item if a given agent $i$ is in $S$ state, we choose at random one of his/her nearest neighbors, say $j$. If such neighbor $j$ is in $P$ state, we generate a random number $r$ in the range $[0,1]$. If $r<\beta$, the agent $i$ changes to state $P$;

\item if a given agent $i$ is in $P$ state, we choose at random one of his/her nearest neighbors, say $k$. If such neighbor $j$ is not in $D$ state, we generate a random number $r$ in the range $[0,1]$. If $r<\sigma$, the agent $i$ changes to state $D$ (spontaneous $P\to D$ transition). On the other hand, if the neighbor $j$ is in $D$ state, we generate a random number $r$ in the range $[0,1]$. If $r<(\delta+\sigma)$, the agent $i$ changes to state $D$ \footnote{This rule takes into account both the spontaneous $P\to D$ transition (probability $\sigma$) and the $P\to D$ transition due to social pressure of $D$ individuals over $P$ ones (probability $\delta$).};

\item if a given agent $i$ is in $D$ state, we generate a random number $r$ in the range $[0,1]$. If $r<\gamma$, the agent $i$ changes to state $A$;

\item if a given agent $i$ is in $A$ state, we generate a random number $r$ in the range $[0,1]$. If $r<\alpha$, the agent $i$ changes to state $D$. Otherwise, if $r>\alpha$, the agent $i$ changes to state $S$.

\end{itemize}

One time step is defined by the visit of all lattice sites. The agents' states were updated syncronously, i.e., we considered parallel updating, as it is standard in probabilistic cellular automata in order to avoid correlations between consecutive steps \cite{mf_keom,roos,religion}. In addition, all results are averaged over $100$ independent simulations.

\begin{figure}[t]
\begin{center}
\vspace{6mm}
\includegraphics[width=0.45\textwidth,angle=0]{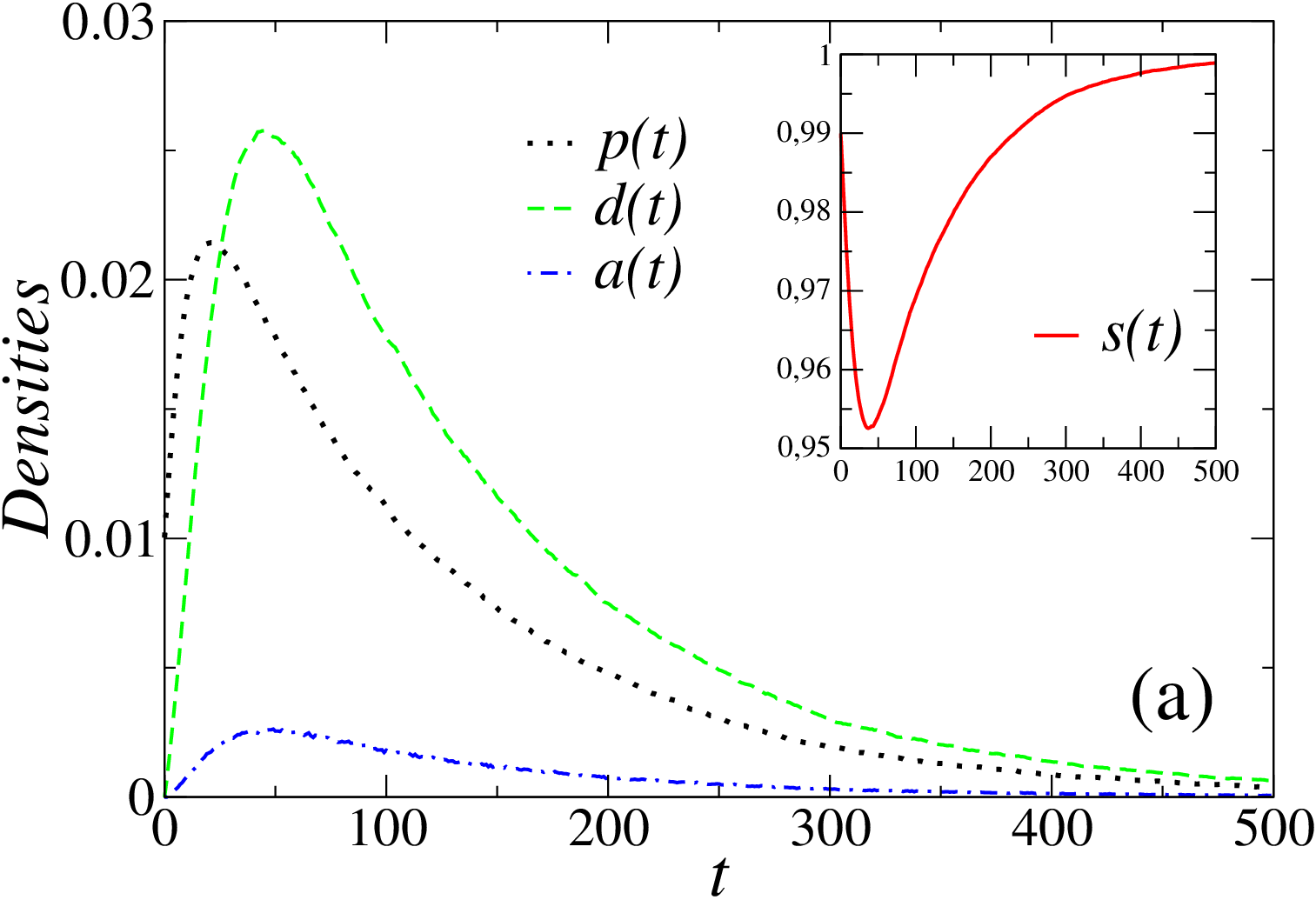}
\hspace{0.3cm}
\includegraphics[width=0.45\textwidth,angle=0]{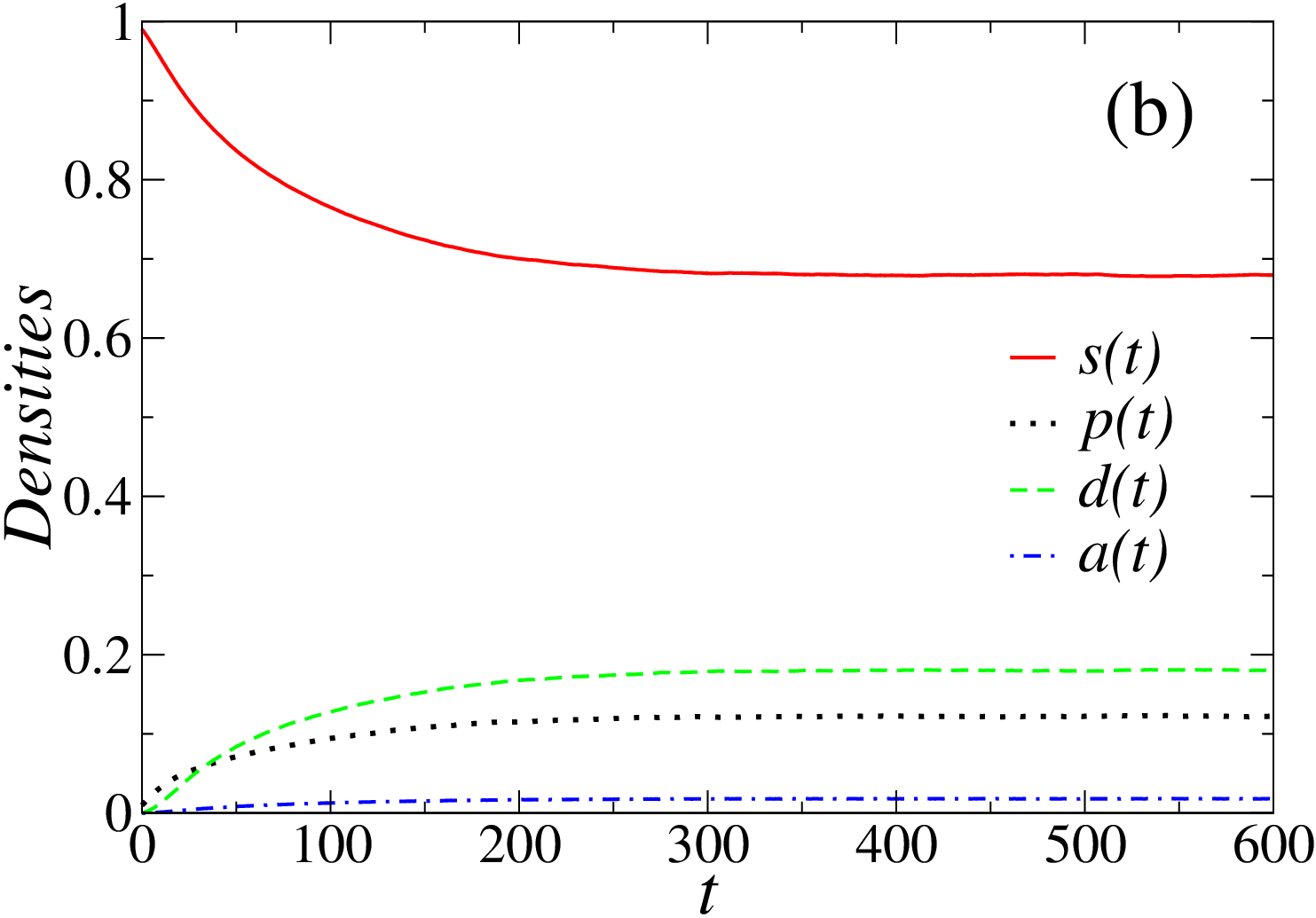}
\\
\vspace{1.0cm}
\includegraphics[width=0.45\textwidth,angle=0]{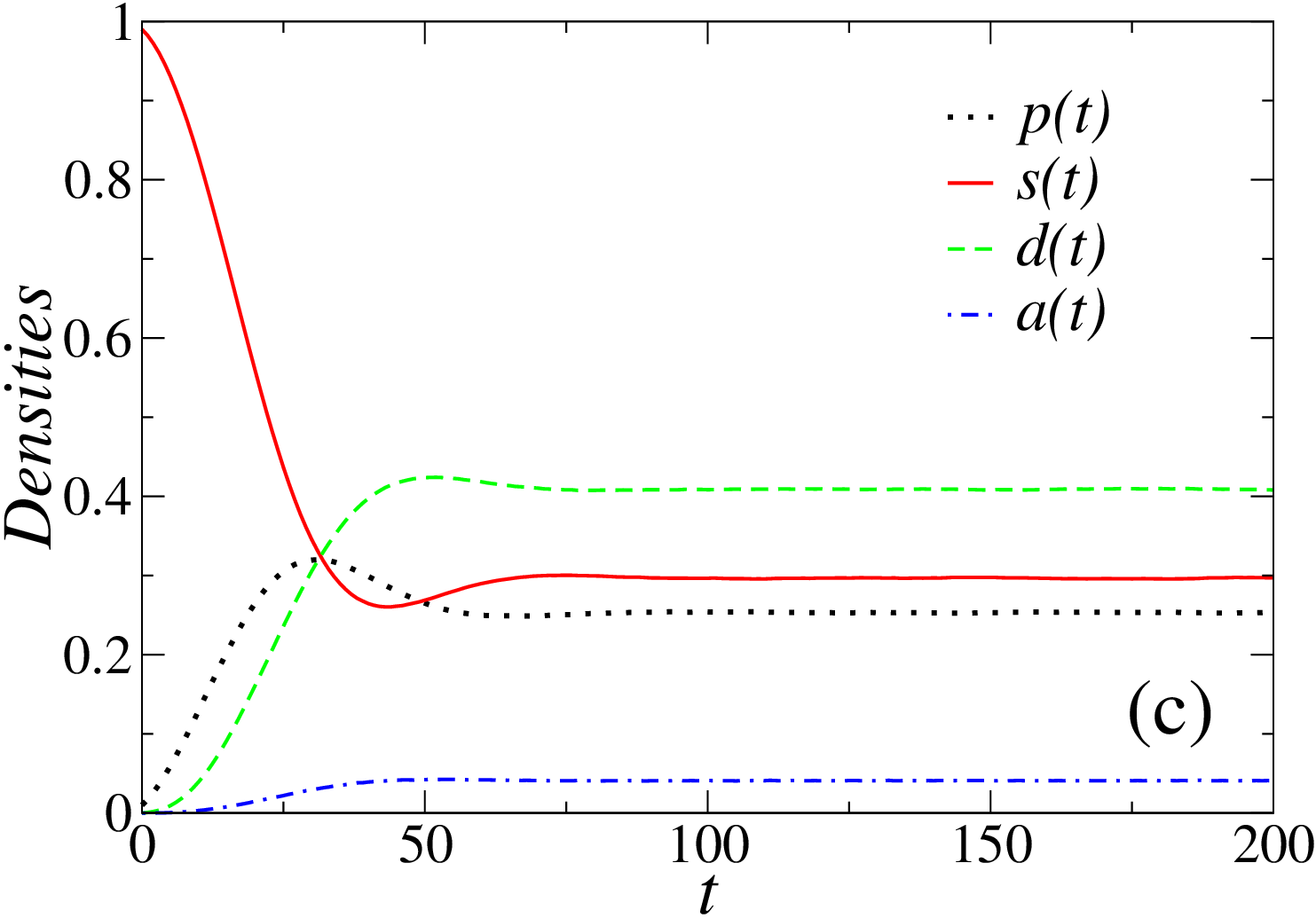}
\hspace{0.3cm}
\includegraphics[width=0.45\textwidth,angle=0]{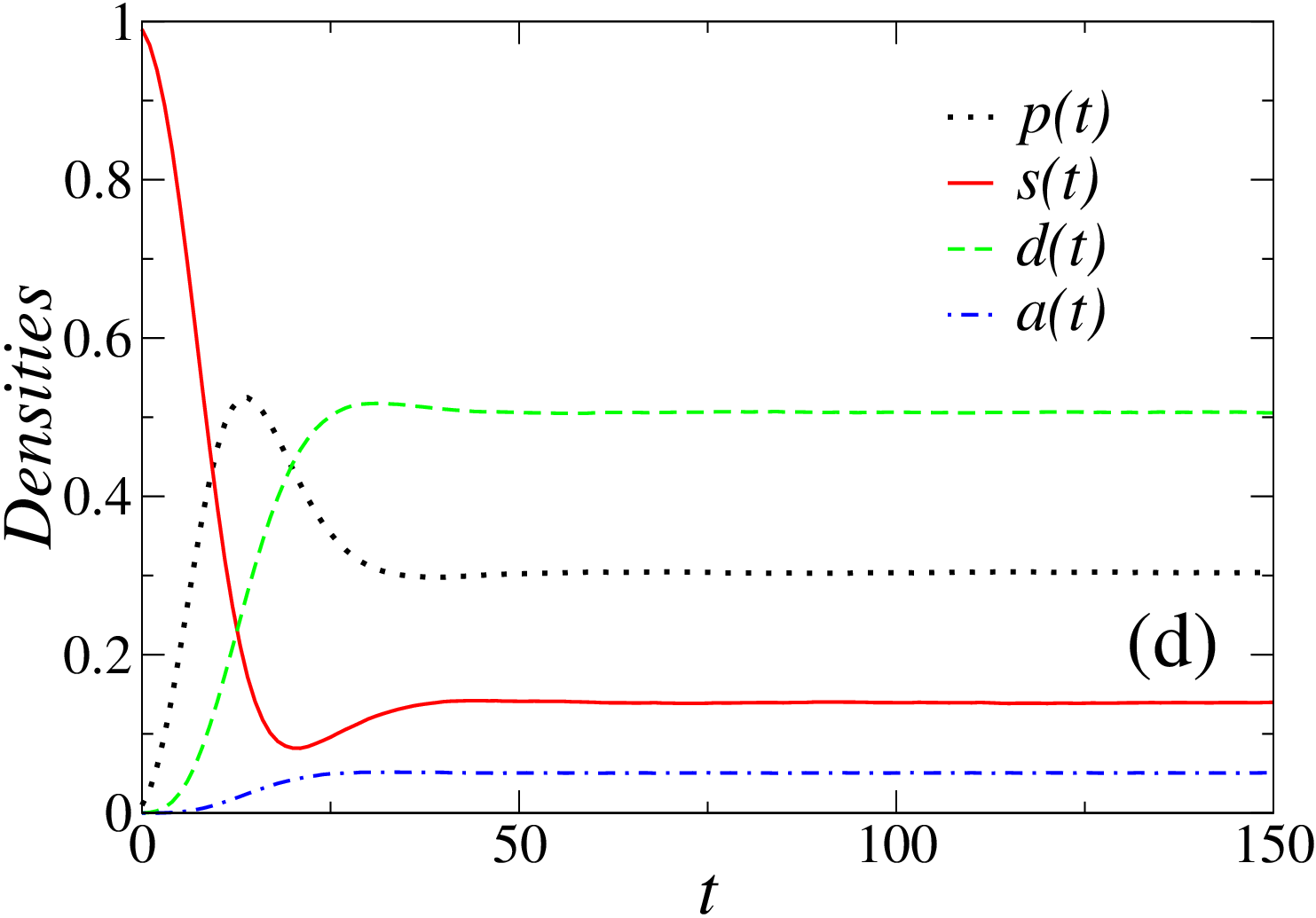}
\end{center}
\caption{Time evolution of the four densities of agents $s(t), p(t), d(t)$ and $a(t)$ for the model defined on a square lattice of linear size $L=100$. The fixed parameters are $\delta=0.05, \alpha=0.30$, $\sigma=0.07$ and $\gamma=0.10$, and we varied the parameter $\beta$: (a) $\beta=0.05$, (b) $\beta=0.08$, (c) $\beta=0.20$, (d) $\beta=0.50$. Results are averaged over $100$ independent simulations.}
\label{fig4}
\end{figure}

Considering the same parameters of the previous subsection, namely $\delta=0.05, \alpha=0.30$ and $\sigma=0.07$, and fixing $\gamma=0.10$, we exhibit in Fig. \ref{fig4} the time evolution of the population on a square lattice of linear size $L=100$ for typical values of $\beta$. We can see a similar behavior observed in the fully-connected case. For example, for sufficient small values of the contagion probability $\beta$ we see a small outbreak for the subpopulations $P, D$ and $A$, but they disapear of the system for long times, and we will observe only $S$ individuals (see Fig. \ref{fig4} (a)). For higher values of $\beta$, we observe the coexistence phase where the four compartments will survive in the stationary states. In general, the times to achieve such stationary states are larger than in the fully-connected network case, as it is usual in agent-based models defined on lattices \cite{pmco}.

\begin{figure}[t]
\begin{center}
\vspace{6mm}
\includegraphics[width=0.65\textwidth,angle=0]{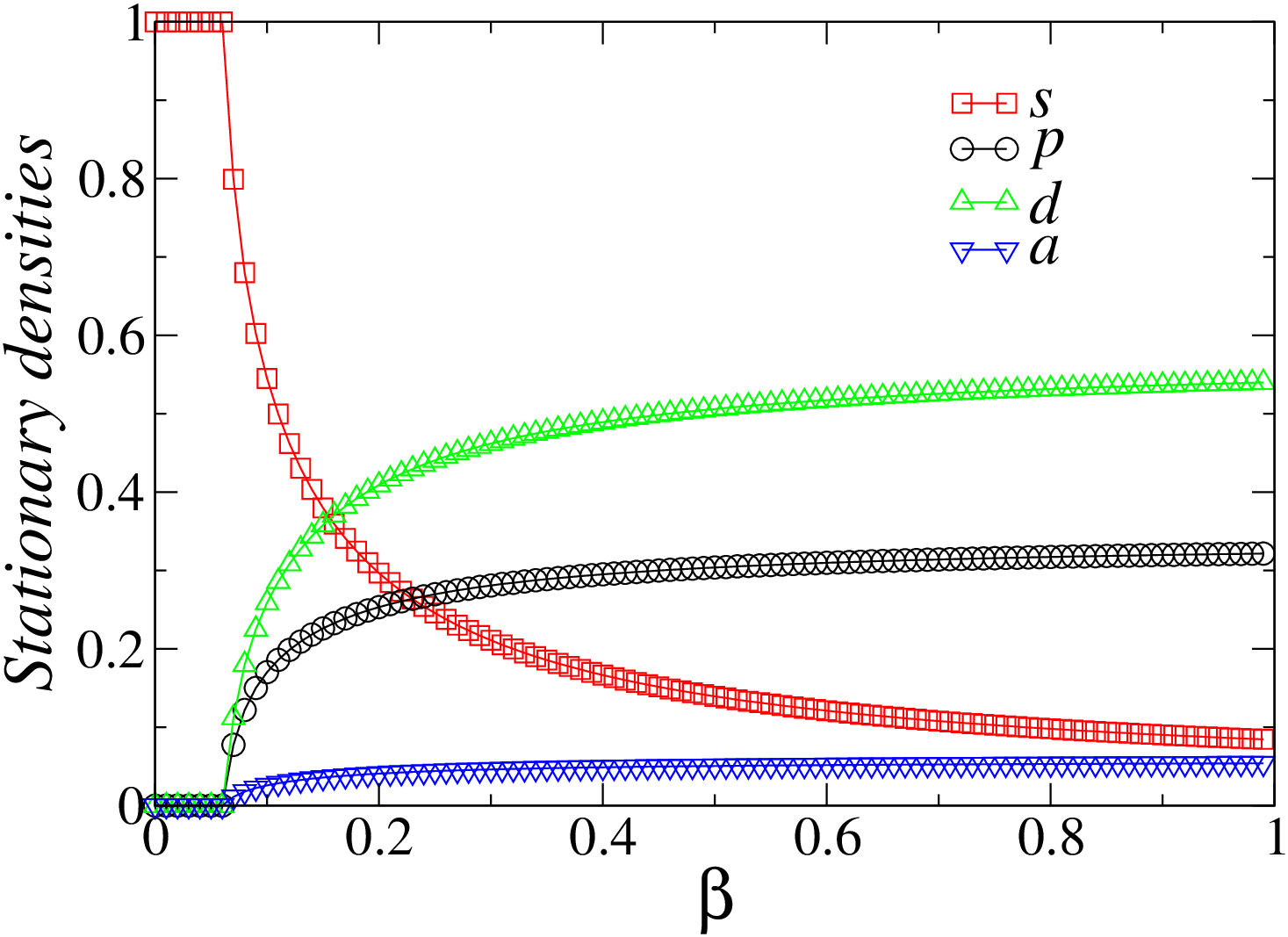}
\end{center}
\caption{Stationary densities $s, p, d$ and $a$ as functions of the contagion probability $\beta$ for $\gamma=0.10, \delta=0.05, \alpha=0.30$ and $\sigma=0.07$ for the model defined on a square lattice of linear size $L=100$. The lines are just guides to the eye. Results are averaged over $100$ independent simulations.}
\label{fig5}
\end{figure}

\begin{figure}[t]
\begin{center}
\vspace{6mm}
\includegraphics[width=0.45\textwidth,angle=0]{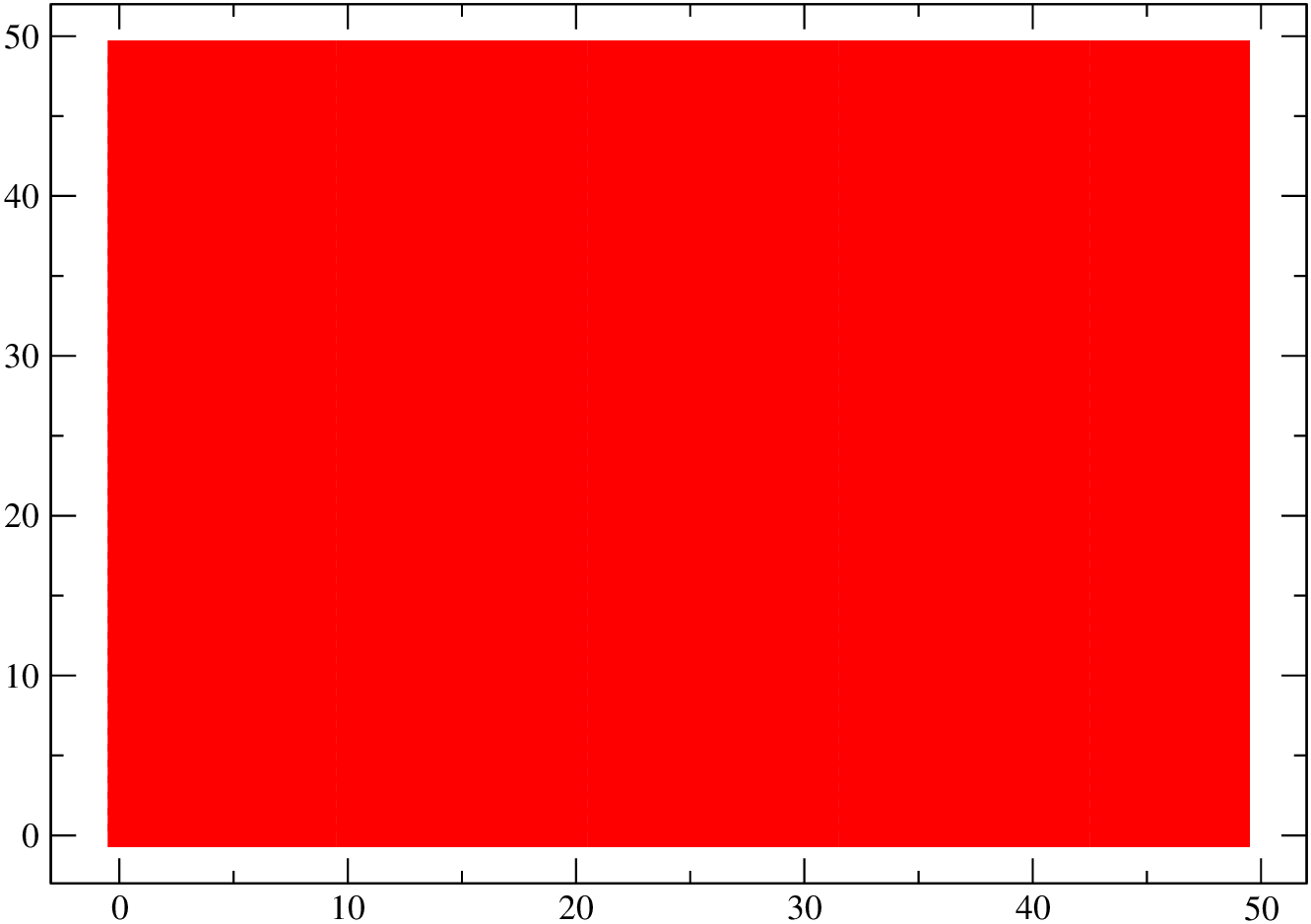}
\hspace{0.5cm}
\includegraphics[width=0.45\textwidth,angle=0]{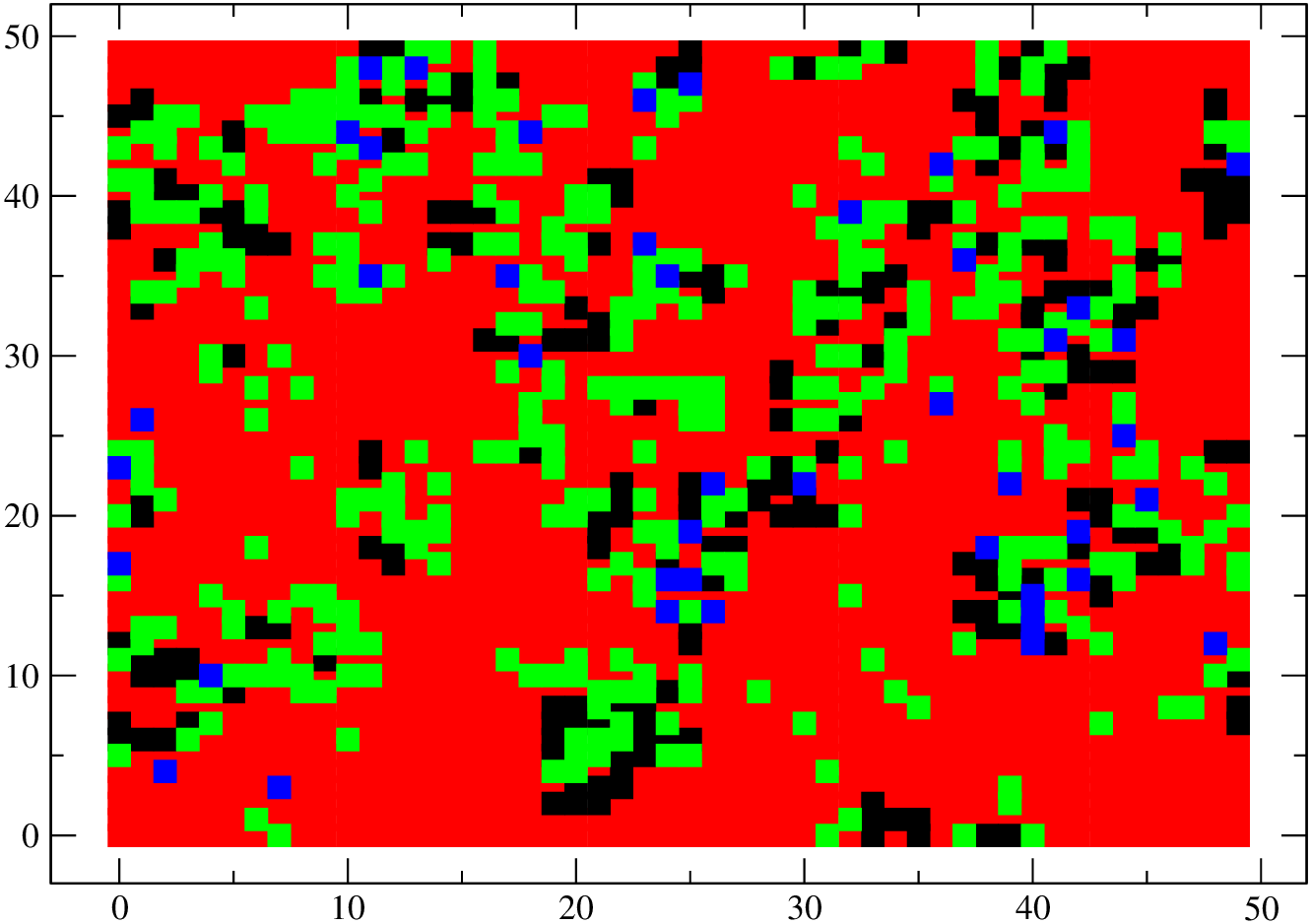}
\\
\vspace{1.0cm}
\includegraphics[width=0.45\textwidth,angle=0]{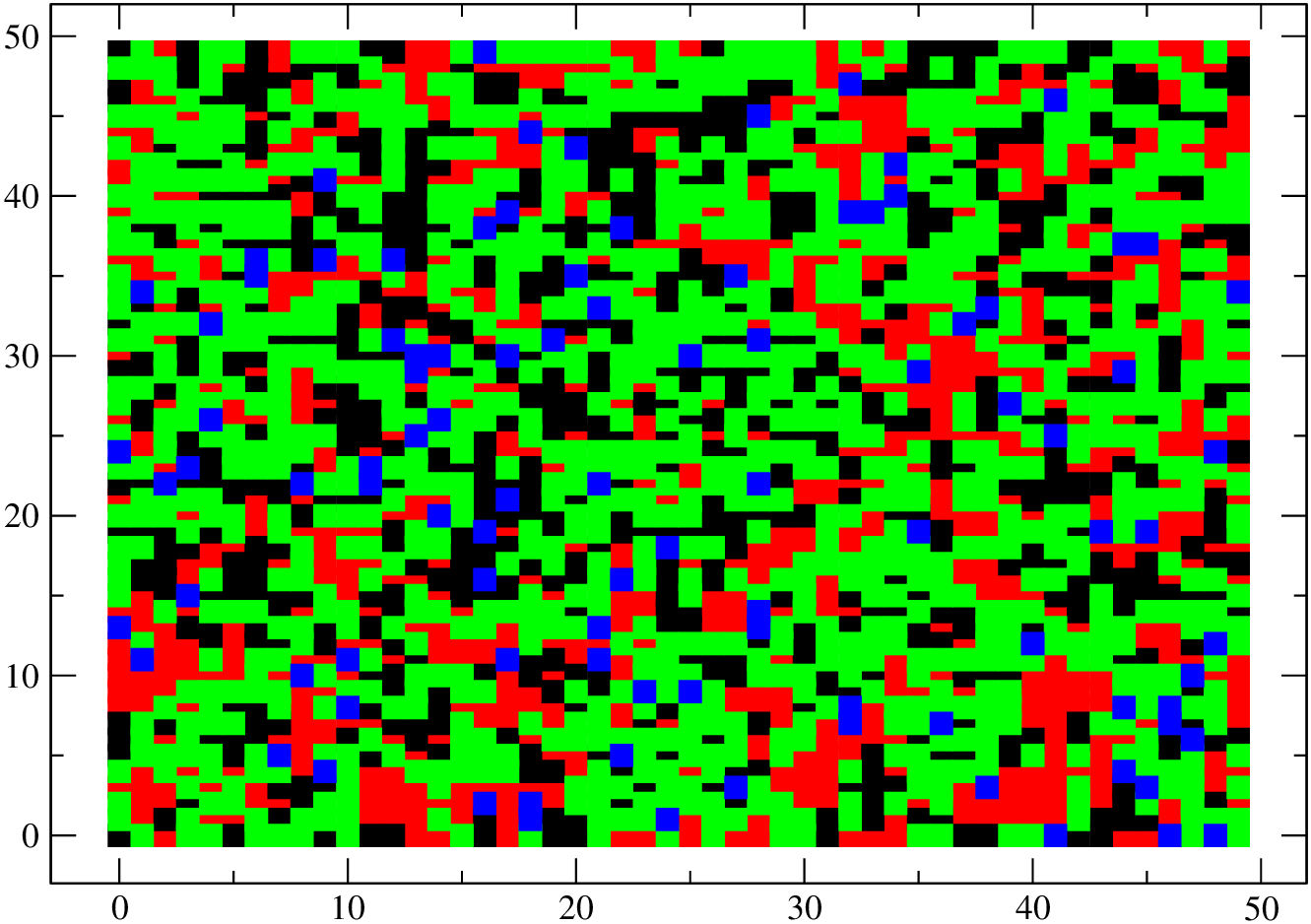}
\hspace{0.5cm}
\includegraphics[width=0.45\textwidth,angle=0]{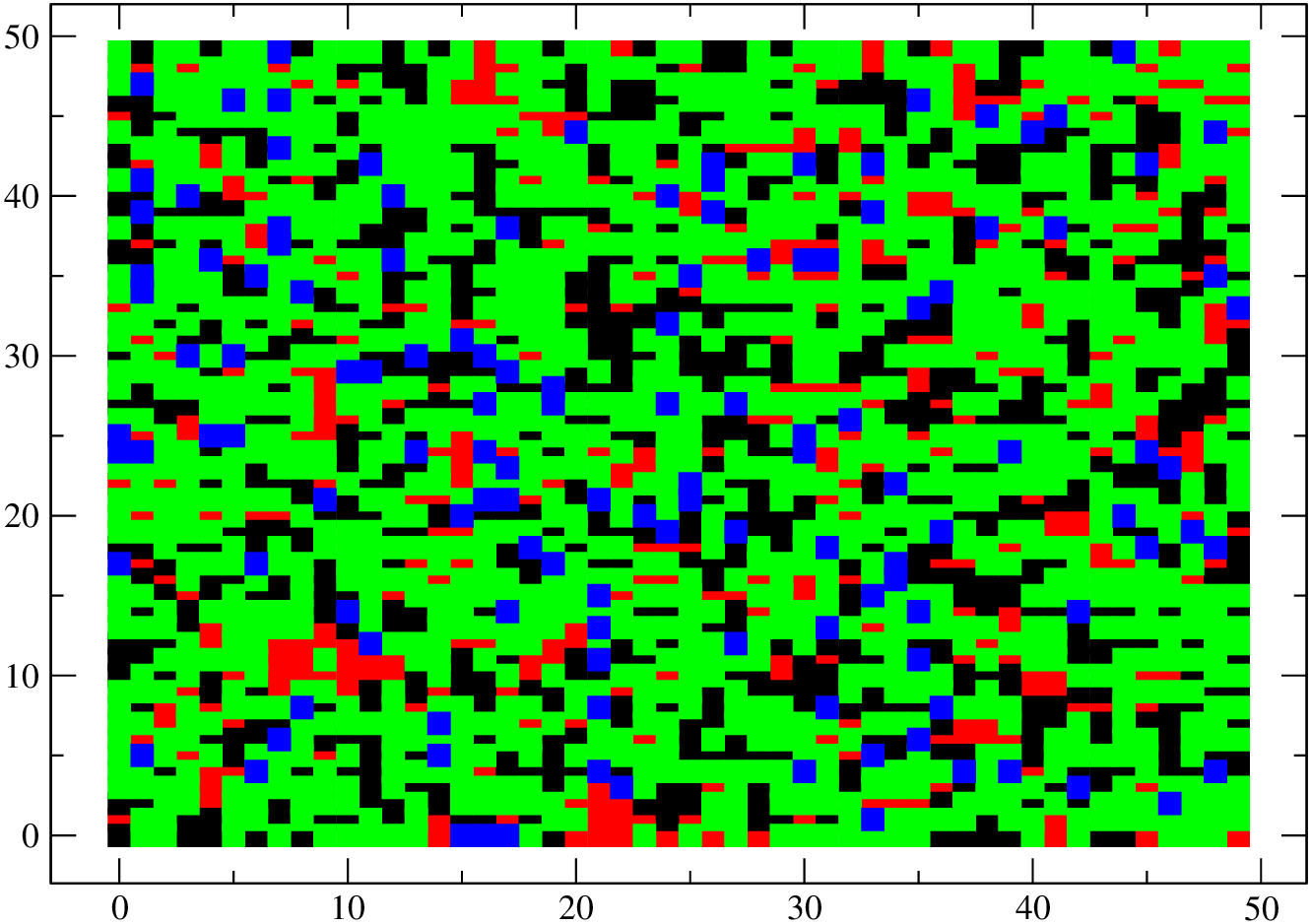}
\end{center}
\caption{Snapshots of the population on a grid of linear size $L=50$ at stationary states. The fixed parameters are $\gamma=0.10, \delta=0.05, \alpha=0.30$ and $\sigma=0.07$, and we varied the contagion parameter $\beta$: (a) $\beta=0.05$, (b) $\beta=0.08$, (c) $\beta=0.20$, (d) $\beta=0.50$. The squares' colors represent distinct subpopulations, namely $S$ (red), $P$ (black), $D$ (green) and $A$ (blue).}
\label{fig6}
\end{figure}

We can also measure the stationary densities $s, p, d$ and $a$ in the simulations in order to compare them with the fully-connect network results. In Fig. \ref{fig5} we exhibit the stationary values of the four subpopulations as functions of the probability $\beta$ for $\gamma=0.10, \delta=0.05, \alpha=0.30$ and $\sigma=0.07$, considering a square lattice with linear size $L=100$. We observe similar behaviors for such quantities in comparison with the analytical calculations presented in the previous subsection. However, as it is usual in lattice models, we observe a distinct location of the critical point $\beta_c$ in comparison with the analytical mean-field calculations presented in the previous subsection. For the same set of parameters, looking for the data for the square lattice case we have $\beta_{c}\approx 0.063$, whereas in the previous subsection we observed $\beta_c=0.07$. This suggest that, in the presence of the lattice, a smaller value of the contagion probability $\beta$ is sufficient to the spreading of drug trafficking in the population. The differences can be pointed to the correlations generated by the lattice structure, that are absent in the fully-connected case.

For better visualization of the lattice and the subpopulations, we plot in Fig. \ref{fig6} some snapshots of the steady states of the model. For this figure we considered the lattice size as $L=50$, the same fixed parameters $\gamma=0.10, \delta=0.05, \alpha=0.30$ and $\sigma=0.07$, and we plot four distinct values of $\beta$. The distinct colors represent the subpopulations $S$ (red), $P$ (black), $D$ (green) and $A$ (blue). For small values of $\beta$ like $\beta=0.05$, we observe the discussed absorbing phase where the subpopulations $P, D$ and $A$ disappear of the population after a long time, and only the Susceptible ($S$) subpopulation survives (upper panel, left side). If we increase $\beta$ to $\beta=0.08$, we observe that the subpopulations $P, D$ and $A$ survive and coexist with the $S$ population, but the Susceptibles are yet the majority in the population (upper panel, right side). For $\beta=0.20$ the drug dealers are the most dominant subpopulation in the grid. Finally, for $\beta=0.50$ the subpopulation $S$ appears in a small number, and most of passive supporters ($P$) turned into drug dealers. Notice also that the drug dealers $D$ (green squares) appear close to passive supporters $P$ (black squares), which remember us the importance of the parameter $\sigma$, that are responsible for the spontaneous $P \to D$ transition. T


\section{Final Remarks}

\qquad In this work we propose a mathematical model in order to study the emergence of drug trafficking in a population. For this purpose, we considered the concept of passive supporters, individuals that do not oppose a drug dealing act (selling of drugs). They go unnoticeable and most of them reject the violent aspect of the drug trafficking action (robberies, confrontation with police force, and so on). The population is divided in four subpopulations, namely susceptibles ($\textbf{S}$), passive supporters ($\textbf{P}$), drug dealers ($\textbf{D}$) and arrested drug dealers ($\textbf{A}$). As it is standard in contagion epidemic-like models, the transitions among such subpopulations are ruled by probabilities.

First, we considered the model in a fully-connected population. Thus, we write the rate equations for the evolutions of the four subpopulations. Analytical results were obtained for the equilibrium solutions of the model. Such solutions, together with the numerical integration of the model's equations, reveal that there are two distinct collective states or phases in the stationary states of the model, depending on the range of parameters: (I) a phase where only the $\textbf{S}$ subpopulation survives and (II) a region where the four subpopulations $\textbf{S, P, D}$ and $\textbf{A}$ coexist. We showed that the emergence of drug dealers in the population is consequence of the presence of passive supporters in the system, and the higher the passive supporter subpopulation, the higher the fraction of drug dealers. We also verified that the police action, modelled by an arrest probability $\gamma$, can be effective to control the spreading of drug dealers. In addition, we also presented in this work results of agent-based Monte Carlo simulations of the model on two-dimensional square lattices. The results show a similar behavior in comparison with the fully-connected case, and we also verified the emergence of drug trafficking as a nonequilibrium phase transition. However, due to the presence of the lattice a smaller value of the contagion probability $\beta$ is sufficient to the spreading of drug trafficking in the population. Finally, the times to the system achieves steady states are higher in comparison with the fully connected case. Emergence of criminality were associated with phase transitions also in other contexts \cite{tax1,tax2,radical,violent}.


\section*{Acknowledgments}

The author acknowledges financial support from the Brazilian scientific funding agencies Conselho Nacional de Desenvolvimento Cient\'ifico e Tecnol\'ogico (CNPq, Grant 308643/2023-2) and Funda\c{c}\~ao de Amparo \`a Pesquisa do Estado do Rio de Janeiro (FAPERJ, Grant 203.217/2017).

\appendix
\section{Analytical calculations: model on a fully-connected network}

In this appendix we will detail some of the analytical calculations considering the model defined in a fully-connected population, presented in section 3.A.

Let us start with the $t\to\infty$ in Eq. (\ref{eq10}). Taking $da/dt = 0$, we obtain 
\begin{equation} \label{app_eq1}
a = \gamma\,d
\end{equation}
\noindent
For Eq. (\ref{eq8}), taking $dp/dt = 0$ we obtain two solutions,
\begin{eqnarray} \label{app_eq2}
p & = & 0 \\ \label{app_eq3}
s & = & \frac{\delta\,d+\sigma}{\beta}
\end{eqnarray}
From Eq. (\ref{eq7}), we obtain
\begin{equation} \label{app_eq4}
a = \frac{\beta}{1-\alpha}\,p\,s
\end{equation}

If the solution $p=0$, Eq. (\ref{app_eq2}), is valid, from Eq. (\ref{app_eq4}) we have $a=0$, which also leads to $d=0$ from Eq. (\ref{app_eq1}). Thus, from the normalization condition, Eq. (\ref{eq11}), we have $s=1$. This solution $(s,p,d,a)=(1,0,0,0)$ defines the absorbing state of the model.

From Eq. (\ref{eq9}), the starionary state gives us
\begin{equation} \label{app_eq5}
p = \frac{(1-\alpha)\,\gamma\,d}{\delta\,d+\sigma}
\end{equation}
\noindent
where we considered $a=\gamma\,d$ from Eq. (\ref{app_eq1}) to simplify the result. Substituting Eqs. (\ref{app_eq1}), (\ref{app_eq3}) and (\ref{app_eq5}) in the normalization condition Eq. (\ref{eq11}), we found a second-order polynomial for $d$, namely $c_{1}d^{2} + c_{2}d + c_{3} = 0$, where
\begin{eqnarray} \label{app_eq6}
c_1 & = & \delta\,[\delta+(1+\gamma)\,\beta] \\ \label{app_eq7}
c_2 & = &  2\,\sigma\,\delta + [(1-\alpha)\,\gamma+(1+\gamma)\,\sigma-\delta]\,\beta \\ \label{app_eq8}
c_3 & = & \sigma\,(\sigma-\beta)
\end{eqnarray}
which solution is
\begin{equation} \label{app_eq9}
d = \frac{c_2}{2\,c_1}\,\left\{-1 \pm \sqrt{1-\frac{4\,c_1\,c_3}{c_2^{2}}}\right\}  
\end{equation}
We verified numerically that the solution of Eq. (\ref{app_eq9}) with the \textit{plus} signal is the relevant one for the problem, leading to $d>0$. We also can verify that we have $d=0$ from Eq. (\ref{app_eq9}) for $\beta=\sigma$, which defines the critical point of the model,
\begin{equation} \label{app_eq10}
\beta_{c} = \sigma
\end{equation}
\noindent
that separates the above-mentioned absorbing phase, for $\beta\leq\beta_{c}$, from the coexistence phase, for $\beta>\beta_{c}$, where the four subpopulations $s, p, d$ and $a$ coexist in the stationary states. The stationary densities $s, p$ and $a$ in such coexistence phase can be found considering the result of Eq. (\ref{app_eq9}) (with the plus signal, as discussed above) in Eqs. (\ref{app_eq3}), (\ref{app_eq5}) and (\ref{app_eq1}), respectively.

The stationary solutions can be also analyzed from the Jacobian matrix point of view. Thus, the stability of the stationary solutions, Eqs. (\ref{eq7}) - (\ref{eq10}) can be analyzed from the eigenvalues of the Jacobian matrix of the system. A given solution is locally asymptotically stable if all eigenvalues of $J$ have negative real parts \cite{rw}. The eigenvalues $\lambda$ can be obtained from $det(J-\lambda\,I)=0$, where $I$ is the identity matrix. The Jacobian matrix of Eqs. (\ref{eq7}) - (\ref{eq10}) is given by
\begin{equation*}
J = 
\begin{bmatrix}
-\beta\,p & -\beta\,s & 0 & 1-\alpha \\
\beta\,p  & \beta\,s-\sigma-\delta\,d   & -\delta\,p   & 0 \\ 
0   & \sigma+\delta\,d    & \delta\,p-\gamma & \alpha \\
0          & 0         & \gamma       & -1
\end{bmatrix}
\end{equation*}

For the absorbing state given by $(s,p,d,a) = (1,0,0,0)$, the eigenvelues of the Jacobian matrix are $\lambda_{1}=0, \lambda_{2}=\beta-\sigma, \lambda_{3}=-\gamma$ and $\lambda_{4}=-1$. Clearly we have $\lambda_{3}<0$ and $\lambda_{4}<0$. Finally, we have $\lambda_{2}<0$ for $\beta<\sigma$, which is the condition for the validity of the absorbing state solution, as discussed in the text. The coexistence solution given by Eqs. (\ref{eq19}) - (\ref{eq22}) is the stable for $\beta>\sigma$. It defines the critical point $\beta_c=\sigma$.


\bibliographystyle{elsarticle-num-names}

\end{document}